\documentclass[aps,pre,groupedaddress,showpacs,preprint]{revtex4}
\usepackage{graphicx}% Include figure files
\usepackage[dvips]{epsfig}% Include figure files
\usepackage{dcolumn}% Align table columns on decimal point
\usepackage{bm}% bold math
\usepackage{amssymb}
\usepackage{amsmath}
\usepackage{subfigure}

\begin{document}
\title{Effects of internal fluctuations on the spreading of Hantavirus}

\author{C. Escudero$^{\dag}$, J. Buceta$^{\dag,\ddag}$, F. J. de
la Rubia$^{\dag}$, and Katja Lindenberg$^{\ddag}$ }

\affiliation{
\dag Departamento de F\'{\i}sica Fundamental,
Universidad Nacional de Educaci\'on a Distancia, C/ Senda del Rey 9,
28040 Madrid, Spain\\
\ddag Department of Chemistry and Biochemistry, and Institute
for Nonlinear Science, University of California San Diego, 9500
Gilman Drive, La Jolla, CA 92093-0340, USA}

\begin{abstract}
We study the spread of Hantavirus over a host population of deer mice
using a population dynamics model. We show that taking
into account the internal fluctuations in the mouse population due
to its discrete character strongly alters the behaviour of the system.
In addition to the familiar transition present in the deterministic
model, the inclusion of internal fluctuations leads to the emergence
of an additional deterministically hidden transition.  We determine
parameter values that lead to maximal propagation of the disease, and
discuss some implications for disease prevention policies.

\end{abstract}

%\keywords{Stochastic Processes, Pattern Formation}
\maketitle

\section{Introduction}
\label{introduction}

Hantavirus epidemics have been studied extensively
in the biological literature following a number of outbreaks in the 
North American Southwest in the 1990's~\cite{mills1}. The host of this
infection is the deer mouse, the most numerous mammal in 
North America. The virus is transmitted among deer mice
via biting and to humans via contact with their excreta.

Recently, the Hantavirus has been receiving increasing attention in
the physical and mathematical literature. A basic population
dynamics model was introduced and solved by Kenkre~\cite{kenkre},
the spatiotemporal patterns of the infection were studied by Abramson and 
Kenkre~\cite{abramson}, Monte Carlo simulations were performed by
Aguirre {\em et al.}~\cite{aguirre}, propagating fronts of
the disease were analyzed by Abramson {\em et al.}~\cite{abramson1}, and the
relation between outbreaks of the disease and seasonal changes was
explored by Buceta {\em et al.}~\cite{buceta}. 
This work has shed light on the mechanisms of propagation of
the disease among mice, and will hopefully help design more 
effective prevention policies.

In this paper we go a step further and analyze the effects of the
\emph{internal} fluctuations on
the propagation of the disease. These fluctuations are inevitable
because the mouse population is discrete and finite, and they may have
profound consequences, as reported by Escudero {\em et al.}~\cite{escudero}
using a generic population dynamics model.

The basic model introduced in~\cite{kenkre} incorporates birth, death,
competition for resources, and infection.  The model reads:
\begin{subequations}
\begin{eqnarray}
\label{meanfielda}
\frac{dM}{dt}&=&(b-c)M-\frac{M^2}{K}, \\
\frac{dM_I}{dt}&=&-cM_I-\frac{M_IM}{K}+a(M-M_I)M_I,
\label{meanfieldb}
\end{eqnarray}
\end{subequations}
where $M$ stands for the total number of mice, $M_I$ for the total
number of infected mice, $b$ is the birth rate coefficient, $c$ the
death rate coefficient, $a$
the infection rate coefficient, and $K$ the carrying
capacity that characterizes the resources available to the
mice and the resulting competition. 
The steady state value of the total mouse population is $M=K(b-c)$.
One can see that there is a transcritical bifurcation
at $K=K_c$, with
\begin{equation}
K_c\equiv \frac{b}{[a(b-c)]}. 
\label{kc}
\end{equation}
When $K<K_c$ the stable point 
\begin{equation}
M=K(b-c),\qquad M_I=0
\label{meanfieldsol1}
\end{equation}
has zero infected mice, while when
$K>K_c$ the stable point includes a 
positive number of infected mice, 
\begin{equation}
M=K(b-c),\qquad M_I=K(b-c)-\frac{b}{a}.
\label{meanfieldsol2}
\end{equation}
The two rate equations can
be thought of as describing two ``reactants,'' $M$ and $M_I$,
undergoing four types of
``reactions'' with rate coefficients $a$, $b$, $c$, and $K^{-1}$
respectively.  One of these conserves the total number of mice (the
infection), while the other three (birth, death, competition) do not.
Note that Eq.~(\ref{meanfielda}) depends only on the latter three, whose
effect on the total population can thus be studied separately from the
issue of infection.  Infected pregnant mice produce Hanta
antibodies that keep their foetus free from the infection.
Consequently, there is no birth term in Eq.~(\ref{meanfieldb}). Also,
there is no recovery term in the model because mice become
chronically infected with the virus.

The analysis of the internal fluctuations in the mouse population
due to the discrete and finite sizes of the populations
requires the generalization of the mean field model to a stochastic
description, e.g., a master equation. In this paper we start with such a
master equation, but at the very outset we outline some
reasonable approximations that lead to a mathematically tractable model.

A full master equation description of the problem would involve 
$P(n,n_I,t)$, the probability distribution function for there to be $n$
total mice and $n_I$ infected mice at time $t$.  We find this full
master equation to be analytically intractable.  We therefore break the
problem up into two parts as follows.  First, we formulate a master
equation for the reduced probability distribution function $P(n,t)$
associated only with the mean field equation~(\ref{meanfielda}).  This
master equation (which is not influenced by the infection) is tractable, as
we shall see.  We then argue that the fluctuations in the
\emph{infected} mouse
population arise from two sources.  One is the dependence on $M$ in
Eq.~(\ref{meanfieldb}) and the fact that this total population
fluctuates.  Having solved the master equation associated with $M$, we
are able to incorporate these fluctuations into the stochastic
description of infected mice. We will show that the effects of these
fluctuations may be profound, especially when the mean mouse population
is not too large, and may lead to unexpected consequences.
These are the new features that we are particularly interested in
exploring.  The other arises from the additional inherent fluctuations
in the number of infected mice due to the fact that this population is
also finite and discrete. These are especially important when
the number of infected mice is small, but we do not include
them explicitly in our equations, again because of tractability
problems. This is not as serious as one might think because we do know
their consequences, which can also be profound (as we have shown 
in~\cite{escudero}): these fluctuations may
cause a small population of infected mice to disappear
entirely.  In other words, if
one is in a regime where the population of infected mice is small
in the absence of these fluctuations, consideration of these fluctuations
might eliminate this population entirely.  Thus, results obtained without
consideration of these fluctuations can be thought of as an upper bound
on the number of infected mice.  At worst one would be overestimating
the presence of infected mice in the regime where the number of infected
mice is in any case small or zero.

In Sec.~\ref{modeltotal} we present the stochastic model for the total
mouse population.  Section~\ref{modelinfected} deals with the stochastic
model for the infected mouse population, and in Sec.~\ref{results} we
discuss the results of the analysis. We summarize our conclusions in
Sec.~\ref{conclusions}.

\section{Stochastic Model for Total Mouse Population}
\label{modeltotal}

The master equation for the total mouse population is easily
written down if we think explicitly of the ``reactions'' contributing to
Eq.~(\ref{meanfielda}).  They are births,
\begin{equation}
M \overset{b}{\rightarrow} M+M,
\end{equation}
deaths,
\begin{equation}
M \overset{c}{\rightarrow} \emptyset,
\end{equation}
and competition for resources,
\begin{equation}
M+M \overset{K^{-1}}{\rightarrow} M.
\end{equation}
The master equation describing these processes is
\begin{eqnarray}
\frac{dP(n,t)}{dt}&=&b\left[(n-1)P(n-1,t)-nP(n,t)\right]
+c\left[(n+1)P(n+1,t)-nP(n,t)\right] \nonumber\\
&&+K^{-1}\left[(n+1)nP(n+1,t)-n(n- 1)P(n,t)\right]. 
\label{masterequation}
\end{eqnarray}
This equation is not tractable as it stands, but it is amenable to
a system size expansion as introduced by van
Kampen~\cite{vankampen,gardiner}.  A system size expansion is
appropriate when the system is ``large'' or, as in our case, the species
under consideration numerous. 
%As we have mentioned, the deer mouse is the
%most numerous mammal in North America. 
It is important to note that
this implies that the steady state solution for $M$ given in
Eqs.~(\ref{meanfieldsol1}) and (\ref{meanfieldsol2}), which we expect
the mean of the stochastic solution to reproduce, must therefore be
``large'', that is, $K$ must be proportional to the
system size.  The ratio 
\begin{equation}
d\equiv \frac{\Omega}{K},
\label{ratio}
\end{equation}
where $\Omega$ is the size of the system, 
must be essentially independent of the system size for this
analysis to be appropriate.
To implement a system size expansion we thus write the third coefficient
on the right of Eq.~(\ref{masterequation}) as $K^{-1} = d/\Omega$. 

The implemention of a system size expansion requires several steps.
First, although $n$ is a discrete variable, we can represent the
discrete changes in $n$ via an infinite series of derivatives in which
$n$ is treated as a continuous variable:
\begin{equation}
f(n \pm 1) = \exp\left(\pm \frac{\partial} {\partial n}\right)
f(n) = \sum_{j=0}^\infty \frac{(\pm 1)^j}{j!} \frac{\partial^j}{\partial
n^j} f(n).
\label{continuous}
\end{equation}
This exact relation allows us to rewrite the master
equation~(\ref{masterequation}) as  
\begin{equation}
\frac{dP(n,t)}{dt}=\left[b \sum_{j=1}^\infty \frac {(-1)^j}{j!}
\frac{\partial^j}{\partial n^j} +c \sum_{j=1}^\infty \frac {1}{j!}
\frac{\partial^j}{\partial n^j}\right] nP(n,t)
+\frac{d}{\Omega} \sum_{j=1}^\infty \frac
{1}{j!}\frac{\partial^j}{\partial n^j} n(n-1)P(n,t).
\end{equation}
Next, one makes the heuristic assumption that one can perform the change
of variables
\begin{equation}
n \to \Omega \phi(t)+\Omega^{1/2}z,
\end{equation}
where $\phi(t)$ is the mean value of the mouse population density and $z$
represents the fluctuations around the mean.
We then define the probability distribution
\begin{equation}
\rho(z,t)=\frac{P(n,t)}{\Omega^{1/2}}.
\label{newprob}
\end{equation}
Applying the standard chain rule 
\begin{equation}
\left( \frac{\partial P(n,t)}{\partial t}\right)_n = \left(
\frac{\partial \rho(z,t)}{\partial t}\right)_z + \left(\frac{\partial
\rho(z,t)}{\partial z}\right)_t \left( \frac{\partial z}{\partial
t}\right)_n 
\end{equation}
together with the relation which follows from the change of
variables relation,
\begin{equation}
\left( \frac{\partial z}{\partial t}\right)_n = -\frac{d\phi(t)}{dt},
\end{equation}
we can (after
multiplying through by $\Omega^{1/2}$) rewrite the master equation in
terms of the new distribution,
\begin{eqnarray}
\label{expansion}
\frac{\partial \rho(z,t)}{\partial t}&-&\Omega^{1/2}\frac{d\phi}{dt}
\frac{\partial \rho(z,t)}{\partial z} \nonumber\\
&&=\left[b \sum_{j=1}^\infty \Omega^{-j/2}\frac {(-1)^j}{j!}
\frac{\partial^j}{\partial z^j} +c \sum_{j=1}^\infty \Omega^{-j/2}
\frac {1}{j!}
\frac{\partial^j}{\partial z^j}\right] \left(\Omega\phi
+\Omega^{1/2}z\right)\rho(z,t)
\nonumber\\
&&+\frac{d}{\Omega} \sum_{j=1}^\infty 
\Omega^{-j/2}\frac{1}{j!}\frac{\partial^j}{\partial
z^j}\left(\Omega\phi+\Omega^{1/2}z\right)
\left(\Omega\phi+\Omega^{1/2}z-1\right) \rho(z,t).
\label{stillexact}
\end{eqnarray}
This equation is still exact.

In the large system size limit there are three divergent terms
in Eq.~(\ref{stillexact})
proportional to $\Omega^{1/2}\partial \rho/\partial z$
that must cancel, that is, we must require that
\begin{equation}
\frac{d\phi}{dt} =(b-c)\phi-d\phi^2.
\end{equation}
Note that this exactly corresponds to Eq.~(\ref{meanfielda}),
that is, $\phi(t)$
is indeed the mean population density.  In the steady state we thus
have 
\begin{equation}
\label{phi}
\phi=\frac{b-c}{d}.
\end{equation}
Using this result in the surviving terms in Eq.~(\ref{stillexact})
in the large $\Omega$ limit
then leads, in the steady state, to the equation
\begin{eqnarray}
\frac{\partial \rho}{\partial t} = 0
&=&(b-c)\frac{\partial (z \rho)}
{\partial z} + \frac{b(b-c)}{d} \frac{\partial^2 \rho}{\partial
z^2}\nonumber\\
&=&\frac{\partial (z \rho)}
{\partial z} + \frac{b}{d} \frac{\partial^2 \rho}{\partial
z^2}.
\label{fokker}
\end{eqnarray}
This is the steady state limit of 
a Fokker-Planck equation for the probability density of the stochastic
variable $z$.  Its solution with natural boundary
conditions at $\pm \infty$ is given by
\begin{equation}
\rho (z)=\left(\frac{d}{2 \pi b}\right)^{1/2}e^{-\frac{d}{2b}z^2},
\end{equation}
a Gaussian distribution centered at zero and of width proportional to
$\sqrt{b/d}$. This in turn implies that the total number $n$ of mice
also has a Gaussian distribution whose mean is the mean number of mice
predicted by the deterministic model and whose width is proportional to 
$\sqrt{Kb}$. 

The internal fluctuations thus do not alter the behavior
of the total number of mice in any dramatic way.  They simply lead to a
Gaussian distribution around the deterministic mean whose width increases
with increasing birth rate and increasing carrying capacity.  However,
as we will see in the following sections, the consequences of
this distribution on the number of infected mice can be unexpected.

\section{Stochastic Model for Infected Mouse Population}
\label{modelinfected}

We now return to the infected mouse population, whose evolution is
described in mean field by Eq.~(\ref{meanfieldb}).  While we are
ignoring the internal fluctuations that arise from the fact that this
population is finite and discrete (as discussed earlier), we do wish to
provide a stochastic description that incorporates the effects of the
fluctuations in the total mouse population $n(t)$.  Our results of the
previous section indicate that in the steady state we can think of
$n(t)$ as a stochastic variable, 
\begin{equation}
n(t)=K(b-c)+\delta n(t) = M+\delta n(t),
\label{Mwithfluct}
\end{equation}
where the fluctuations $\delta n(t)$ have zero mean and are generated
from the Ornstein-Uhlenbeck stochastic differential equation
\begin{equation}
\frac{d\delta n}{dt} = - (b-c) \delta n + \sqrt{2Kb(b-c)}\xi(t).
\end{equation}
Here $\xi(t)$ is zero-centered $\delta$-correlated Gaussian noise
of unit intensity, $\langle \xi(t)\xi(t')\rangle=\delta(t-t')$. In the
stationary state the correlation function of the fluctuations is then
\begin{equation}
\langle \delta n(t) \delta n(t') \rangle = Kb e^{-(b-c)|t-t'|}.
\end{equation}
We include these fluctuations in Eq.~(\ref{meanfieldb}) by replacing $n$
with $K(b-c)+\delta n$.
The resulting stochastic differential equation reads
\begin{equation}
\label{stochastic}
\frac{dn_I}{dt}=\left[aK(b-c)-b\right]n_I
-an_I^2+\frac{a-K^{-1}}{\sqrt{b-c}} n_I \zeta(t),
\end{equation}
where $\zeta(t)$ is an Orstein-Uhlenbeck process with zero mean
and correlation function
\begin{equation}
\langle \zeta(t) \zeta(t') \rangle = Kb(b-c) e^{-(b-c)|t-t'|}.
\label{basicinf}
\end{equation}
This is our basic stochastic equation for the infected population.
The intensity of these fluctuations is determined by the width of the
total mouse population distribution.  The correlation time
$\tau_c=(b-c)^{-1}$ is a measure of the time it takes a total population
diminished by fluctuations to recover.

Because $\zeta(t)$ is a ``colored noise'' with finite correlation time,
the exact solution of the problem (\ref{stochastic})-(\ref{basicinf})
is not known. In particular, there is no exact Fokker-Planck equation
for the probability distribution function $P(n_I,t)$ that the number of
infected mice is $n_I$ at time $t$.
A number of approximate Fokker-Planck equation
schemes can be found in the literature~\cite{katja}, some of which have
the virtue of becoming exact in both the limits    
$\tau_c \rightarrow 0$ and $\tau_c \rightarrow \infty$.
Since these theories lead to a qualitatively similar panorama of
possibilities, we apply
the simplest of these theories, developed by Fox~\cite{fox1,fox2}.
The resulting effective Fokker-Planck equation is
\begin{equation}
\frac{\partial}{\partial t}P(n_I,t)=-\frac{\partial}{\partial n_I}
G(n_I)P(n_I,t)
+\frac{\partial}{\partial n_I}g(n_I)\frac{\partial}{\partial n_I}
g(n_I)D(n_I)P(n_I,t),
\end{equation}
where
\begin{equation}
G(n_I)=[aK(b-c)-b]n_I-an_I^2,
\end{equation}

\begin{equation}
g(n_I)=\frac{a-K^{-1}}{\sqrt{b-c}} n_I,
\end{equation}
and
\begin{equation}
D(n_I)=\frac{Kb(b-c)}{b-c+an_I}.
\end{equation}
The stationary solution of this equation is
\begin{eqnarray}
P(n_I)&=&N \left(1+ \frac{a}{b-c}n_I \right)
n_I^{\left(\displaystyle -1+\frac{\displaystyle (b-c)K[a(b-c)K-b]}
{\displaystyle b(aK-1)^2}\right)} \nonumber\\ \nonumber\\
&&\times \exp\left[ - \frac {K a^2n_I^2}{2b(aK-1)^2} + \frac
{aK\left[aK(b-c)-2b +c\right]n_I}{b(aK-1)^2}\right],
\label{probdistr}
\end{eqnarray}
where $N$ is the normalization constant.  In the next section
we analyze and comment on the interesting features of this solution.

\section{Results for Infected Mouse Population}
\label{results}

\begin{figure}
\begin{center}
\subfigure[$~b=0.8$, $c=0.5$]{\includegraphics[width=7cm]{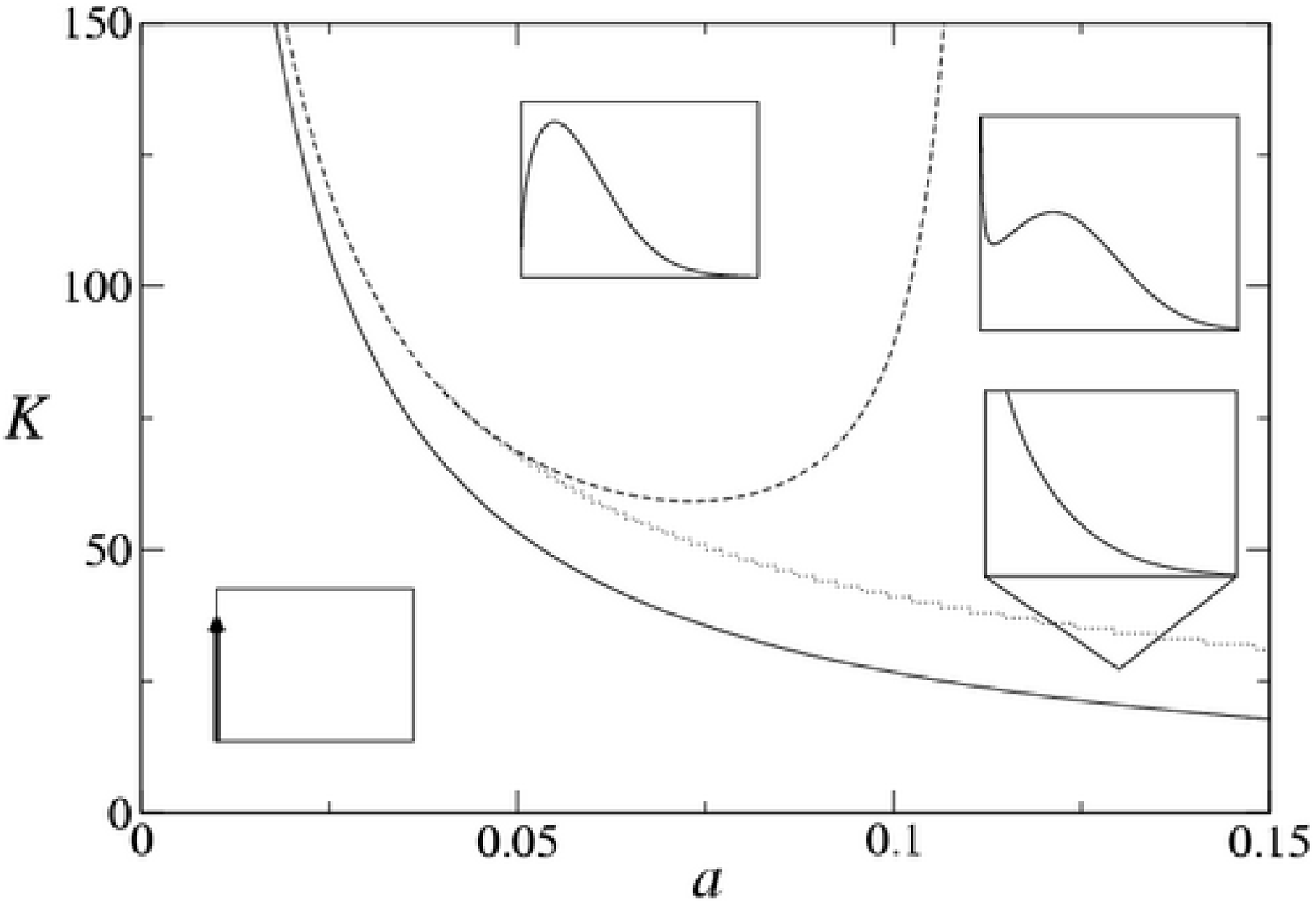}}
\subfigure[$~b=0.5$, $c=0.2$]{\includegraphics[width=7cm]{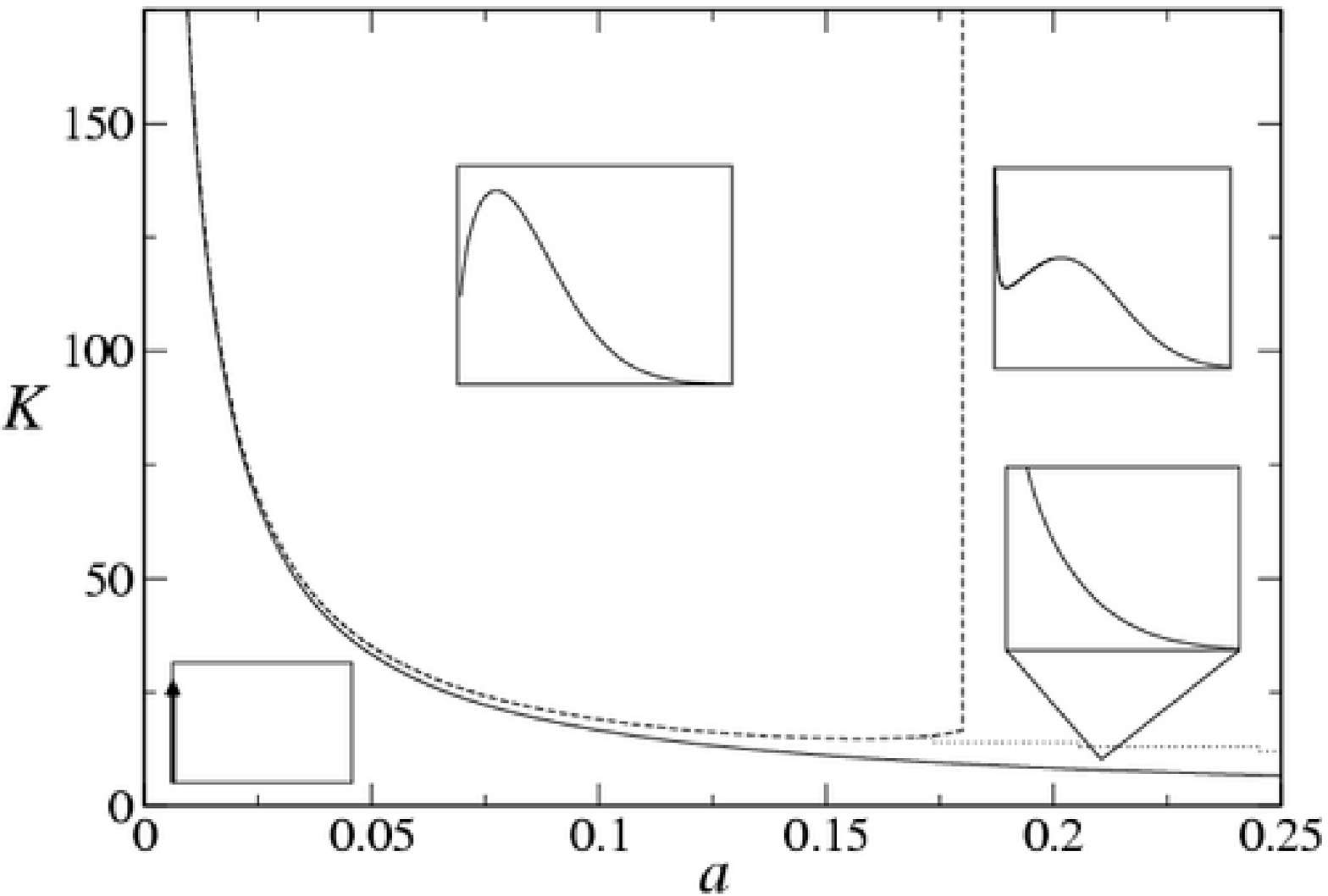}}
\end{center}
\caption{Phase diagrams for the infected mouse population in $(K,a)$
space, described in detail in the text.
\label{figshape}}
\end{figure}

The first point to note is that the mean number of mice,
\begin{equation}
M_I = \langle n_I \rangle = \int_0^\infty dn_I n_I P(n_I),
\end{equation}
is exactly as predicted in the mean field theory, cf.
Eqs.~(\ref{meanfieldsol1})-(\ref{meanfieldsol2}).
However, here there are distributions of infected mouse populations
underlying this mean, and
our interest lies in the different shapes of these
distributions in different parameter regimes, and in the additional
information beyond the mean contained in these distributions.
The distributions, described in detail below, are sketched in
Fig.~\ref{figshape}, where we present the phase diagrams of the system
in $(K,a)$ space for fixed $b$ and $c$.  The values of $b$ and $c$ have
been chosen to match those used in earlier Monte Carlo
simulations~\cite{aguirre} and/or to make most evident the different
behaviors that are observed in the system.

Various phase boundary lines are shown in Fig.~\ref{figshape}. The solid
curves in both panels are the curves $K=K_c$. When $K<K_c$ the
probability distribution (\ref{probdistr}) can not be normalized because
it has a nonintegrable singularity at $n_I=0$. Since this is a fixed
point of the dynamics, the probability distribution must be interpreted
as a $\delta$ function centered at zero~\cite{gardiner}. The insets at
the lower left of each panel are a schematic of this behavior.  When $K$
crosses the $K_c$ curve there is still a divergence at $n_I=0$, but
the probability distribution becomes integrable and hence normalizable. 
The most probable value for the infected population is still zero, but
nonzero values now have a finite probability and the mean value of
infected mice is positive.  The lower right hand insets in both panels
are sketches of this behavior, which persists as a function of
increasing $K$ until the carrying capacity reaches a second critical
value,
\begin{equation}
K_c^* = \frac{2b}{b[2a-(b-c)]+\sqrt{b(b-c)[b^2-c(4a+b)]}}.
\label{kc2}
\end{equation}
When $K$ crosses the curve $K=K_c^*$ the divergence in the distribution
(\ref{probdistr}) disappears and the most probable number of infected
mice moves to finite values. The curve has a divergent asymptote at
$a=0$, but its behavior as a function of $a$ otherwise depends on the
other parameters.  If $b<2c$ then $K_c^*$ also diverges at
$a=a_c \equiv(b-c)^2/b$.  This is the
situation in panel (a) of Fig.~\ref{figshape}.  On the other hand, if
$b>2c$ then $K_c^*$ is complex when $a>a_c\equiv b(b-c)/4c$,
thus producing the abrupt vertical boundary seen in
panel (b).  This second case corresponds to the
parameters $b$ and $c$ in the Monte Carlo simulations of Aguirre
{\em et al.}~\cite{aguirre}.  In either case, within the region enclosed by 
the $K_c^*$ curve (dashed curve in the figure) the probability
distribution goes to zero at the origin and has a maximum at a finite
value of $n_I$, as shown in the upper left sketches in both panels. 
Both the average number of infected mice and the most probable
number of infected mice are now positive.  Note that we have labeled the
rightmost value of $a$ on the dashed curve as $a_c$, whether it is an
asymptote as in panel (a) or the abrupt ending point of the curve as
in panel (b).

There is an additional transition curve, more subtle than the other
two, drawn as the dotted curves in both panels in Fig.~\ref{figshape}.
We denote this transition curve as $K_c^{**}$.
On the upper right hand in both panels is the sketch of the
probability distribution in this region. Here the
probability distribution diverges at zero, 
but another maximum develops at a finite number of infected mice.
This maximum is found as the finite
positive root of the derivative condition $dP(n_I)/dn_I=0$.  
The dotted curves in the phase diagrams indicate the location
of this transition. 

\begin{figure}
\begin{center}
\subfigure[$a=10^{-2}<a_c$]{\includegraphics[width=7cm]{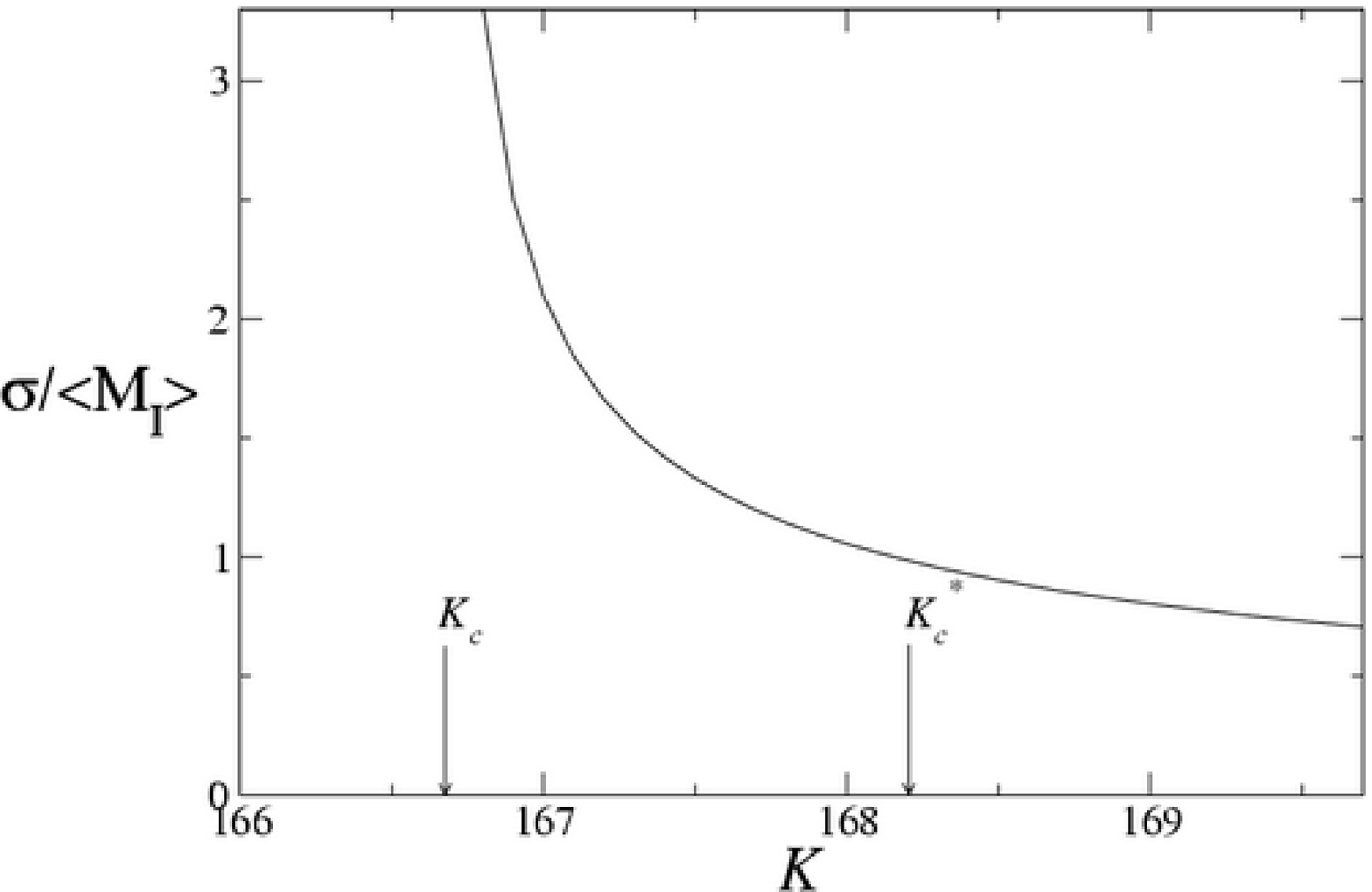}}
\subfigure[$a>a_c$]{\includegraphics[width=7cm]{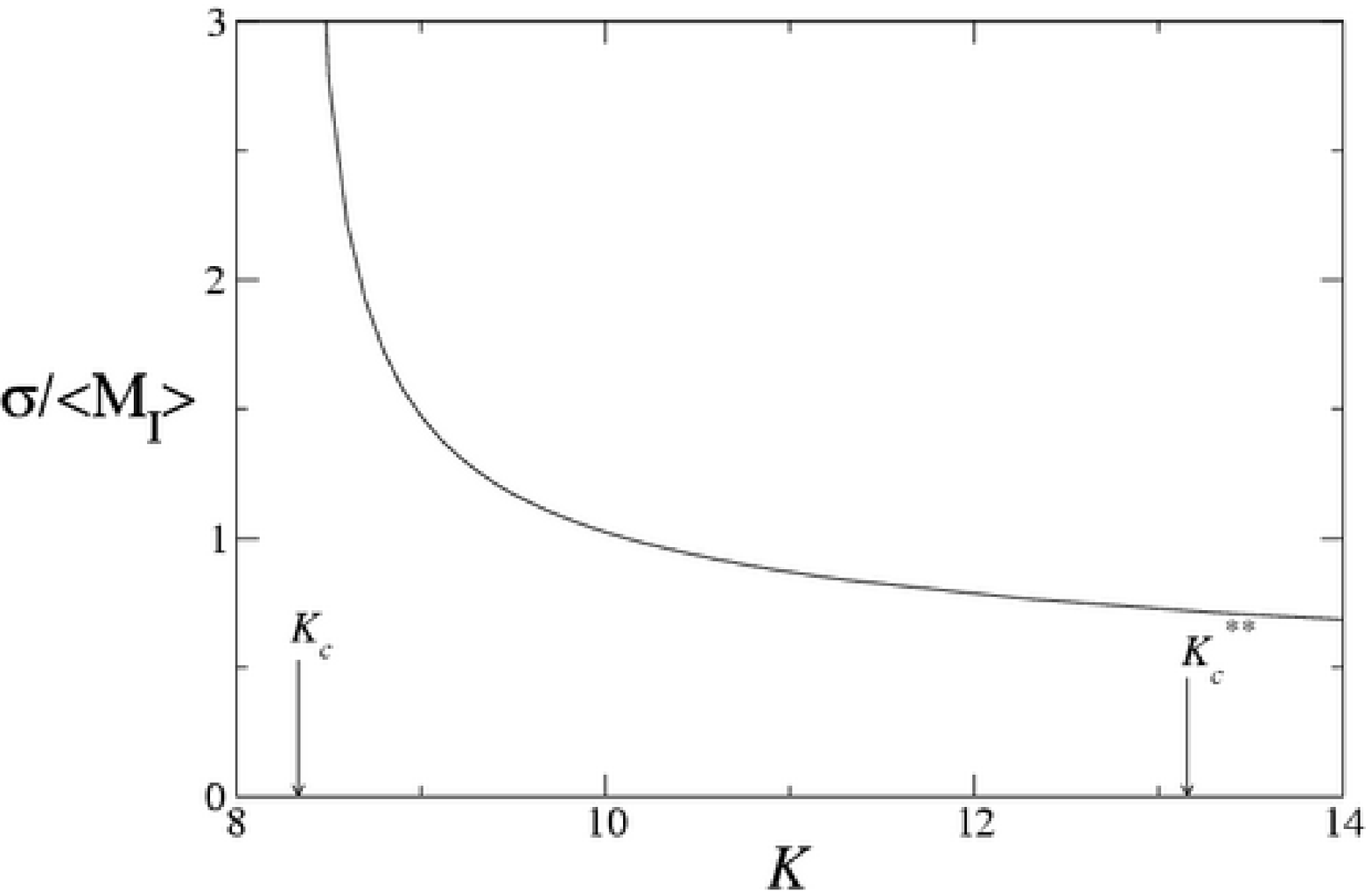}}
\end{center}
\caption{Ratio of the dispersion to the mean number of infected
mice for $b=0.5$ and $c=0.2$.
\label{dispersion}}
\end{figure}

A number of points about these results deserve special highlighting.
%\begin{itemize}
%\item
The particular behavior just described for large $K$ and $a$
(divergence at the origin and also another maximum) is entirely
due to the fact that the internal fluctuations are \emph{colored}.  The
correlation time of these fluctuations is $(b-c)^{-1}$, and the color has
arisen naturally and not as an additional assumption.
%\item
It is interesting to note that
the Monte Carlo simulation results of Aguirre {\em et al.}~\cite{aguirre}
exhibit a number of features that might be related to the results that we
have derived here.  One is that in their simulations the number of
infected mice as a function of $K$ jumps discontinuously
from zero to a finite number (whereas the mean field value does not). 
They note that an explanation for this result lies in the fact that the
number of mice is discrete and that one whole mouse is needed to obtain
a result different from zero (although their jump is much
greater than unity).  In our continuous language, the
behavior they observe might reflect
the abrupt transition between the $\delta$-function
distribution (or the one with a maximum at the origin) to the one with a
zero probability density of no infected mice as $K$ increases.  Their
simulations use the value $a=10^{-2}$. 
%\item
To support this argument further, we have plotted in
Fig.~\ref{dispersion} the
ratio of the dispersion to the mean for these parameter values.  The
dispersion is of the order of the mean and, near the transition value
$K_c$, the ratio actually diverges.  We have elsewhere pursued the
argument that
a possible criterion for the likely extinction of a species is precisely
that the dispersion be of the same size as the mean~\cite{escudero}.  The
substantial width of the distribution might make itself apparent in
a simulation through the high likelihood of absence of the infected species.
For comparison, we have also plotted the ratio of the dispersion to the
mean for a value of $a>a_c$.  The fluctuations are now decidedly
smaller, even though we are in a regime of far fewer infected mice on
average (as indicated by the values of $K$).  It would be interesting to
see whether the size of the jump in the infected mouse population at the
transition would decrease in a Monte Carlo simulation with $a>a_c$.
%\end{itemize}

Finally, we comment on three last points. One concerns the validity of
the system size expansion.  The total mean number of mice in the population
is $K(b-c)$, and the system size expansion is valid if this number is
in some sense sufficiently large (the expansion is valid if
the neglected terms are small).  While we have not explicitly checked
the validity, in most of the regimes under discussion the number of
infected mice is an order of magnitude greater than unity.  Our second
point is to stress that the fluctuations that lead to the distributions of
infected mice are entirely due to the discrete and finite character of
the \emph{total} number of mice.  And yet, while the ratio of the width
of the distribution of total mice to the mean number of total mice,
$\sigma_M/M$, is small in most of the phase diagram
($\sigma_M/M = \sqrt{(b/K)}/(b-c)$), the width of the distribution
induced in the number of infected mice is relatively large (of
$O(\geq 1)$) in most of the diagram.  The third point is a reminder that
this theory has not included the fluctuations caused directly by the
fact that the number of infected mice is discrete and finite.  These
fluctuations would further broaden the distributions.

\section{Conclusions}
\label{conclusions}
We have considered the effect of the internal fluctuations in the total
mouse population on the number of infected mice.
Although these fluctuations cause no dramatic effects in the total
mouse population, yielding a Gaussian distribution of relatively small
variance for a sufficiently large population, these fluctuations have a
rather strong effect on the distribution of infected mice.  Because the
fluctuations are not ``direct'' but instead appear indirectly through
the coupling between infected and uninfected mice, the fluctuations
necessarily and naturally appear as colored in the equations
that describe the evolution of infected mice.  This
leads to a variety of effects beyond those that would be caused by
simple white noise~\cite{horsthemke,katja}.
The mean infected population in this
model is exactly that predicted in mean field.
However, while
the mean field model predicts one critical value of the carrying
capacity parameter ($K_c$) such that below this value there is
no infection and
above this value there is, the stochastic model leads to three
critical values ($K_c$, $K_c^*$ and $K_c^{**}$). The first, which occurs
at the same critical value as
that of the deterministic model, here corresponds to a transition between
a state with no infected mice to an intermediate state in which
the most probable state is still one with no infected mice but with a
finite probability of infection. The second describes a transition
between this intermediate state and the outbreak state, where the
probability distribution that there is no infection goes to zero.
The intermediate state ($K_c<K<K_c^*$) displays different
behaviors depending on the parameter values. In particular, in some
parameter ranges the intermediate state has very few
infected mice.  We argued that the inclusion of the internal
fluctuations in the infected mouse population (which was not considered
due to analytic difficulties) would probably lead to extinction of
this small number of infected mice.  This then means that the 
effective transition between non-epidemic and epidemic states may occur at
$K_c^*$ rather than at $K_c$.  We also identified another transition
curve, $K_c^{**}$, beyond which the probability diverges at zero but
where another maximum develops at a finite number of infected mice.

These features may be useful in the design of more effective
prevention policies. For instance, an increase in the
effective annihilation rate of the mice (by either
increasing the death rate or decreasing the birth rate or both) might
help because it increases the relative size of the region in parameter
space in which the infected mouse population distribution has a
divergence at the origin (the state of no infected mice).
The most effective strategy for the control of Hantavirus outbreaks is
the reduction of the carrying capacity $K$ so as to cross
from one regime to another with a higher probability of no infected mice.

\begin{acknowledgments}
C. Escudero is grateful to the Department of Chemistry and Biochemistry
of the University of California, San Diego for its hospitality.
J. Buceta wishes to acknowledge support by the La Jolla Interfaces in
Science Interdisciplinary Program funded through the generosity of the
Burroughs Wellcome Fund.  This
work has been partially supported by the Engineering Research Program of
the Office of Basic Energy Sciences at the U. S. Department of Energy
under Grant No. DE-FG03-86ER13606, by the Ministerio de Educaci\'{o}n
y Cultura (Spain) through grant No. AP2001-2598, 
and by the Ministerio de Ciencia y Tecnolog\'{\i}a (Spain),
Project No. BFM2001-0291. 
\end{acknowledgments}

\end{document}